# *Todos estos edificios se hacen de diverso modo que en Europa*: Estudio arqueoastronómico de las iglesias jesuíticas de Chiquitos

Alejandro Gangui

IAFE, UBA-CONICET, Argentina

El estudio de la disposición espacial de las iglesias cristianas ha sido de gran interés desde la Antigüedad tardía e inicios de la Alta Edad Media, y ha recibido un nuevo impulso en la literatura reciente al reconocerse que la orientación de sus ejes principales representa un rasgo clave de su arquitectura. Desde las comunidades cristianas más tempranas, la orientación de la iglesia permitía que los fieles rezaran mirando hacia el oriente, en la dirección del Sol naciente. Varios autores antiguos se preocuparon por señalarlo en sus escritos. Y en particular, previo al concilio de Nicea (325 d.C.), ya las Constituciones Apostólicas, en su libro segundo, sección séptima, indicaban: «Y sea, primero, el edificio alargado, con su cabecera hacia el este...».

Aquí presentamos un análisis detallado de la orientación espacial de las iglesias históricas ubicadas en los pueblos jesuíticos de la Chiquitanía, en el oriente boliviano. En nuestro trabajo de campo, hemos medido *in situ* las características principales de ocho iglesias actualmente en pie y las ruinas de una novena construcción de la que se conserva sólo la torre exenta del campanario e indicios de sus muros laterales. En todos los casos, efectuamos un relevamiento detallado del paisaje que rodea a cada iglesia tratando de hallar algún patrón común, eventualmente astronómico, que explique sus orientaciones. A nuestra lista agregamos, además, las medidas de orientación de una décima iglesia (Santo Corazón de Jesús) a partir del trabajo con planos y mapas satelitales. Complementamos nuestros datos con un estudio cultural e histórico detallado de las características de los pueblos misionales.

A diferencia de las iglesias de los pueblos guaraníes de la provincia histórica de Paraquaria, en donde se destacan las orientaciones meridianas en la dirección norte-sur, en el caso de las iglesias jesuíticas chiquitanas la mitad de las construcciones medidas muestra orientaciones que entran dentro del rango solar, entre los acimuts extremos del movimiento anual del Sol al cruzar el horizonte local, con tres iglesias orientadas equinoccialmente con alta precisión. En nuestro trabajo analizamos las razones de estas orientaciones y comentamos brevemente la posible relevancia que los efectos de iluminación –muy buscados en la arquitectura barroca por su relación con la luz natural– podrían haber tenido para los artífices de estas iglesias, que representan verdaderas catedrales escondidas en la selva virgen tropical.

**Introducción**

La orientación de las iglesias cristianas históricas ha ofrecido durante muchos años un interesante campo de estudio académico en el contexto de la Astronomía Cultural. En los últimos tiempos este tema recibió un nuevo impulso en la literatura, ya que se reconoció que la orientación representa una característica clave de su arquitectura. Con base en los textos de los primeros escritores y apologetas cristianos, sabemos que el eje de simetría y los ábsides de las iglesias antiguas debían estar en una dirección bien definida, de modo que el sacerdote mirara hacia el este durante los oficios (McCluskey 2015; Gangui et al. 2016).

Con excepciones notables, como las catedrales de San Juan de Letrán o San Pedro en Roma, que apuntan sus ábsides hacia poniente, la mayoría de las iglesias ya estudiadas en la literatura se dirigen *ad orientem*, con orientaciones que se ubican aproximadamente dentro del rango solar, es decir, con acimuts que caen entre los solsticios de invierno y verano (por ejemplo, la iglesia madre jesuita de Il Gesù en Roma que apunta hacia el horizonte oriental). En estos estudios se evidencia un notable agrupamiento alrededor de los equinoccios y, en ocasiones, también alrededor de los solsticios (González-García 2015, 272).

En este trabajo nos concentramos en las iglesias misionales jesuíticas de Sudamérica que, por más de dos siglos, fueron las construcciones más representativas en el proceso de evangelización cristiana en el continente hasta la expulsión de la orden en 1767. Nuestro objetivo principal es discernir posibles patrones de orientación en las estructuras estudiadas y evaluar si estas orientaciones están relacionadas con la ubicación del Sol u otros cuerpos celestes al cruzar el horizonte local, lo que podría arrojar información importante relacionada con su construcción.

Ya ha habido extensos y detallados estudios históricos y culturales de los pueblos misionales y de sus construcciones más emblemáticas en toda esta región. Sin embargo, la orientación de las iglesias en estos pueblos no había sido objeto de un estudio profundo hasta hace poco tiempo (Giménez Benítez et al. 2018). Además, los trabajos de campo arqueoastronómicos, que consideran las características urbanísticas y los escritos y crónicas de los propios misioneros, solo se han realizado recientemente en los territorios de las misiones guaraníes que, como sabemos, hoy se distribuyen en una gran región que se extiende por tres países: el noreste de Argentina, el sur de Brasil y Paraguay.

Con la idea de continuar y complementar el estudio de las iglesias jesuíticas de la provincia histórica del Paraguay, aquí presentamos un análisis arqueoastronómico de las orientaciones de las iglesias jesuíticas de la Chiquitanía. El trabajo de campo consistió en la medición precisa de las orientaciones de las ocho iglesias que aún existen en la región, y de las ruinas de una novena iglesia, San Juan Bautista de Taperas, de la que se conserva en pie solo la monumental torre-campanario. Mediante el empleo de mapas satelitales y la reconstrucción de la topografía local a partir de un modelo digital del terreno (Kosowsky 2017) pudimos incluir los datos de orientación de una décima iglesia, Santo Corazón de Jesús, ubicada en una zona un poco más apartada de nuestro recorrido planeado (Figura 1). Por último, todas nuestras mediciones de alturas de los horizontes fueron luego corroboradas con los mismos modelos digitales del terreno.

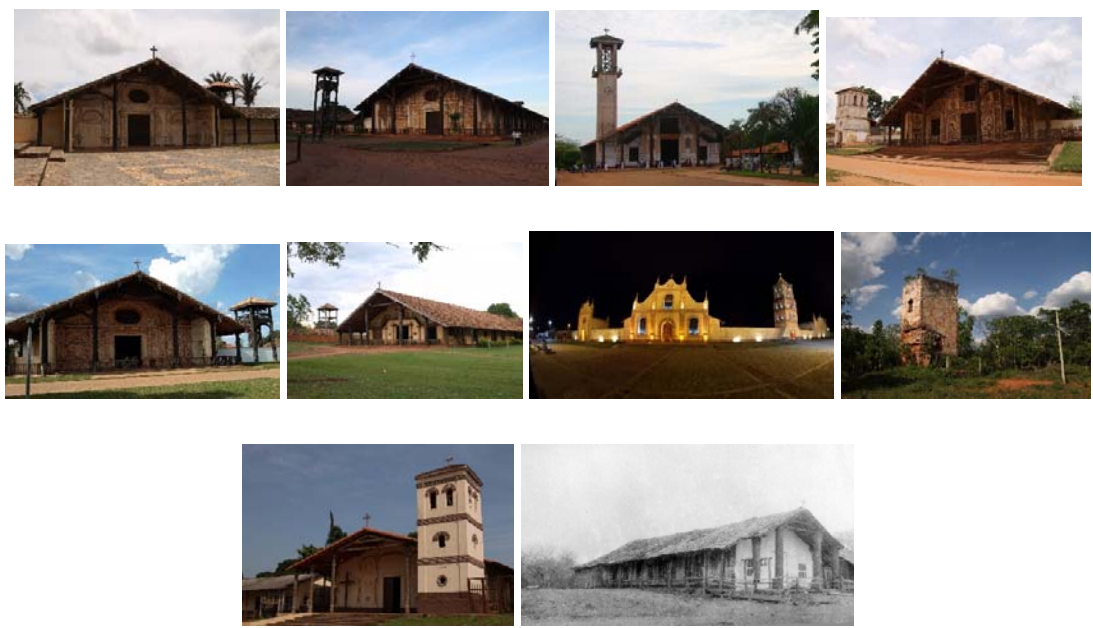

Figura 1. Las diez iglesias jesuíticas de la provincia de Chiquitos medidas en este estudio. De izquierda a derecha y de arriba abajo: San Francisco Javier, Nuestra Señora de la Inmaculada Concepción, San Ignacio de Velasco, San Miguel de Velasco, San Rafael de Velasco, Santa Ana de Velasco, San José de Chiquitos, San Juan Bautista de Taperas (se muestran las ruinas de su campanario), Santiago de Chiquitos y Santo Corazón de Jesús (en una foto de época del coronel Cuéllar extraída de Querejazu 1995, foto 355).

En lo que sigue presentaremos los elementos históricos y culturales que han caracterizado el asentamiento de los jesuitas en el oriente boliviano (Querejazu 1995). Luego repasaremos la fundación de los pueblos, su urbanización y el establecimiento de los edificios religiosos, especialmente de las iglesias que, como veremos, representan verdaderas catedrales escondidas en la selva virgen tropical. Presentaremos entonces los datos de nuestras mediciones y su análisis. Trataremos así de comprender parte de los motivos que guiaron a sus constructores a emplazarlas y orientarlas de la manera en que lo hicieron.

Veremos que a pesar del pequeño número de construcciones analizadas es posible extraer algunas conclusiones interesantes. Una de ellas es que, a diferencia de las iglesias de la provincia de Paraquaria, en donde la gran mayoría de las orientaciones se ubicaba en el cuadrante sur (Ruiz Martínez-Cañavate 2017, 168; Giménez Benítez et al. 2018), en el caso de las iglesias chiquitanas la orientación de la mitad de los templos medidos se ubica dentro del rango solar (Gangui 2020). Más aún, tres de estas construcciones están orientadas equinoccialmente con un error menor a los 2°. Estas son las iglesias de Concepción y San José, cuyos altares apuntan hacia el este, y San Francisco Javier, cuya cabecera se dirige hacia el oeste. Para terminar, discutiremos brevemente la relevancia que los efectos de iluminación podrían haber tenido en el interior de los templos. Estimamos que este es un elemento importante para tener en cuenta en estudios futuros, ya que podría revelar parte de las razones que expliquen el patrón de orientaciones de las construcciones de la orden Jesuita en la región.

**Los jesuitas de Chiquitos**

La Compañía de Jesús fue creada en 1540 y apenas 28 años más tarde ya encontramos jesuitas llevando a cabo su misión en el Virreinato del Perú. Poco tiempo después realizan fundaciones en lugares estratégicos de la Audiencia de Charcas, como en Potosí, en La Paz y, en 1587, en Santa Cruz de la Sierra. Inicialmente los jesuitas concentraron su misión evangelizadora con los chiriguanos de la región cordillerana, pero a fines del siglo XVII una serie de circunstancias los llevó a hacerse cargo de la provincia de Chiquitos, en el actual oriente boliviano. Los chiquitanos, "siempre en guerra" según Charlevoix (1913 [1757], 167), se habían convertido en una seria amenaza para la gobernación de la ciudad de Santa Cruz. Pero al mismo tiempo, los indios eran presa fácil de los cruceños traficantes de esclavos, quienes los capturaban y vendían en regiones cercanas y hasta en el Perú. Además, las incursiones de los *bandeirantes paulistas* mostraban la debilidad de la gobernación de esa ciudad en la zona limítrofe de Chiquitos.

Quien fuera gobernador de Santa Cruz entre 1686 y 1691, Agustín de Arce y de la Concha, estaba convencido de que el problema de los chiquitanos podría resolverse por medio de la evangelización, y para ello recurrió a la Compañía de Jesús. En esos días el padre José de Arce, jesuita natural de La Palma, en las canarias, ya se hallaba en el Colegio de Tarija, formándose para convertir a la fe de Cristo a los feroces chiriguanos. Fue entonces que tomó conocimiento del pedido del gobernador. Inmediatamente se ofreció a trabajar en Chiquitos y, con la venia del padre provincial Gregorio Orosco, que en ese entonces visitaba Tarija, cambió su destino de la cordillera por el de Santa Cruz y, tiempo más tarde, por el de la Chiquitanía (Hoffmann 1979, 169).

Pero una vez en tierra cruceña comenzaron los problemas para el padre Arce. La presencia de misioneros en Santa Cruz no era vista con buenos ojos por los locales, pues ponía en peligro el redituable negocio de venta de indios chiquitos como esclavos. A pesar de una cierta oposición de los lugareños y de la inminencia de la época de lluvias, que harían más dificultoso su recorrido, el padre Arce inició su viaje el 2 de septiembre de 1691, acompañado de otro hermano de su orden y de dos guías indígenas. No sin dificultad arribó al primer pueblo chiquitano, y lo hizo en un momento oportuno, pues los indígenas venían soportando los estragos causados por una peste, muy probablemente una epidemia de viruela, y la presencia del misionero y su ayuda fueron muy bien recibidos (Fernández 1895 [1726], I, Cap. IV, V.I-85).

Es entonces que los aborígenes piden al padre Arce que se quede con ellos y el jesuita interpreta este evento como una señal divina (Fernández 1895 [1726], I, Cap. IV, V.I-87). Así lo hace y en el último día del año 1691, fiesta de San Silvestre, Arce decide fundar la primera reducción de Chiquitos, que más tarde sería dedicada a San Francisco Javier (Parejas Moreno 1995, 273). Con mayores o menores dificultades y con el correr de los años y la intervención de muchos sacerdotes jesuitas, el territorio chiquitano se fue poblando de una multitud de reducciones para los indígenas (Figura 2). Bajo una organización de disciplina rígida, característica de la Compañía de Jesús, en estos pueblos los indios encontraron seguridad para sus familias y bienestar material.

Por muy diversas razones, como el emplazamiento en territorios adversos y la falta de recursos, como el agua, o por los continuos ataques de los paulistas o de parcialidades enemigas de los indios reducidos, la misión de San Francisco Javier y la mayoría de las que poblaron la Chiquitanía debieron trasladarse más de una vez a lo largo de décadas, incluso luego de la expulsión de los jesuitas. En muchos casos las fechas de fundación y los sitios originales son inciertos, pero, excluyendo San Juan Bautista de Taperas, las nueve misiones que aún hoy permanecen en pie son "pueblos vivos" que conservan muchas de las tradiciones de hace siglos y, especialmente, el fervor religioso de la época de los jesuitas.

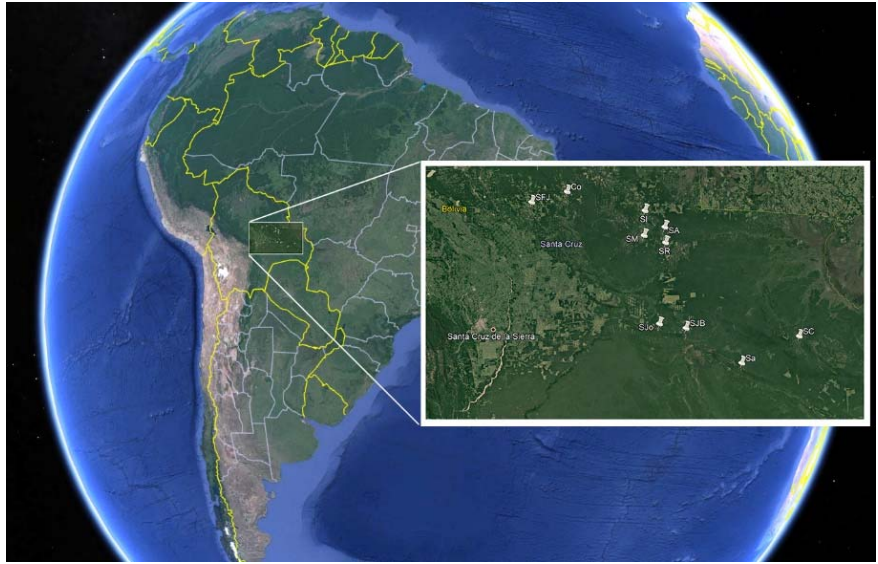

Figura 2. Las diez misiones jesuíticas de la provincia de Chiquitos, en el oriente boliviano. En el mapa se muestra la ubicación de las iglesias de San Francisco Javier (SFJ), Nuestra Señora de la Inmaculada Concepción (Co), San Ignacio (SI), San Miguel (SM), Santa Ana (SA), San Rafael (SR), San José (SJo), San Juan Bautista (SJB), Santiago (Sa) y Santo Corazón (SC).

**La urbanización de los pueblos**

A poco de comenzada la ocupación del territorio chiquitano, los jesuitas aprendieron de los indios acerca de los sitios más adecuados para la instalación de las nuevas reducciones. Para establecer los pueblos los misioneros eligieron "unos lugares llanos a las extremidades de las lomas, a modo de istmos, a fin de que por tres partes fuesen separados por valles de las demás colinas y dispusiesen por ende de un horizonte abierto y les proporcionase un aire libre y sano" (Grondona 1942, 92).

En muchas ocasiones los padres aprovecharon la topografía del lugar y las honduras de los valles cercanos, controlando los niveles y gradientes de una red de pequeños arroyos en zonas de bajíos (Suárez Salas 1995, 418). Esto les permitió formar algunas lagunas artificiales que abastecieron a las reducciones con agua potable permanente para el consumo familiar y la producción agropecuaria.

La organización del espacio urbano en los pueblos misionales tenía su centro en la plaza y, junto a esta, se erigía el templo, el colegio-residencia y el cementerio. Estas construcciones, junto a la torre-campanario, formaban el núcleo religioso esencial de la configuración del espacio (Gutiérrez da Costa y Gutiérrez Viñuales 1995, 338).

Para instalar los pueblos el padre Knogler señala: "se tala el monte en un ámbito suficientemente amplio y se queman las maderas y la maleza; así se traza un cuadrángulo con una plaza grande en el medio, de trescientos o cuatrocientos metros de largo y otros tantos de ancho; en torno a la plaza se levantan las casas de los indios [...]. Tres costados de la plaza son ocupados por estas casas, el cuarto queda reservado para la iglesia, el cementerio, y el colegio donde viven los misioneros y se encuentran los talleres y la escuela" (Knogler 1768 [1979], Cap. II, 147).

La monumentalidad de muchas de las misiones era notoria. Incluso décadas después del extrañamiento de los jesuitas de Chiquitos, ordenada por decreto del rey Carlos III, los viajeros quedaban sorprendidos por la dimensión y la calidad de las plazas (Figura 3). Alcides D'Orbigny, por ejemplo, quien visitó San José de Chiquitos en 1831, decía que su plaza "es enorme, decorada en el centro con una cruz de piedra rodeada de palmeras" (D'Orbigny 1945, tomo III, 1180).

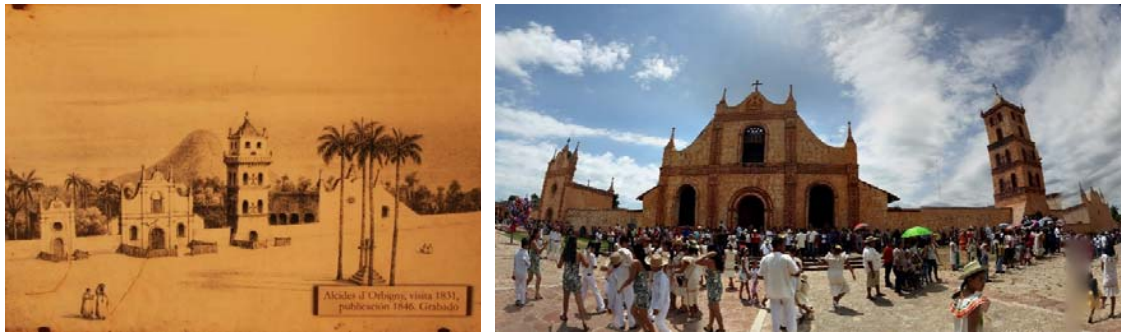

Figura 3. Vista de la plaza de San José de Chiquitos en un dibujo de D'Orbigny de su visita en 1831. El cerro Turubó se destaca en el fondo. A su lado, el frente del complejo religioso de San José en la actualidad durante un día de fiesta. San José se destaca de las otras iglesias patrimoniales por su sólida construcción en mampostería y ladrillo, y por su perfil típicamente barroco europeizante.

Desde un punto de vista cultural, el urbanismo chiquitano incorpora las manifestaciones del espíritu barroco. La plaza es vista como el escenario donde transcurre la vida del pueblo, el punto de donde se parte para ir a trabajar, el sitio para socializar o para los juegos, el lugar donde se desarrollan las fiestas cívicas y religiosas. La plaza como representación del Teatro del Mundo, donde la deidad – único espectador privilegiado – contempla la escena, es una característica de la idea barroca de los usos urbanos (Gutiérrez da Costa y Gutiérrez Viñuales 1995, 338).

Por su parte, las imponentes iglesias – especialmente aquellas con su fachada retablo y su balcón-capilla abierta, que tiende a sacralizar el espacio externo – y las demás construcciones religiosas forman el telón de fondo de una obra en continua representación y son elementos característicos de la propuesta del urbanismo barroco. Lo mismo podemos decir de la búsqueda de integración de ciertos espacios verdes con la selva circundante –por ejemplo, la quinta con frutales y la huerta, ubicadas detrás del núcleo religioso esencial. Estos y otros, son aspectos que muestran la expresión del dominio y el control del paisaje tan propios del barroco.

**La orientación de las misiones**

A partir del plano de la Figura 4 podemos ver que la iglesia se construía con su eje paralelo a la avenida principal de ingreso al pueblo, la que partía desde la capilla Betania (un pequeño oratorio abierto). Pero este eje no estaba alineado con dicha avenida. Es una novedad en el urbanismo misional de Chiquitos que el eje de la plaza y del pueblo coinciden con el eje del patio del colegio, y marca una diferencia importante con el urbanismo de los pueblos de la provincia de Paraquaria, donde es el eje de la iglesia el que coincide con el de la plaza (Roth 1995, 488; Giménez Benítez et al. 2018). La recta continuación de esta avenida pasaba por el centro de la plaza, e iba a dar al primer patio (el principal o patio jesuita), lindante al templo, donde se hallaba un cuadrante o reloj de Sol. Sin embargo, no quedan registros o crónicas escritas por los propios padres fundadores que indiquen la orientación espacial de esta gran estructura (el pueblo), ya sea con el paisaje circundante o con cualquier otro referencial.

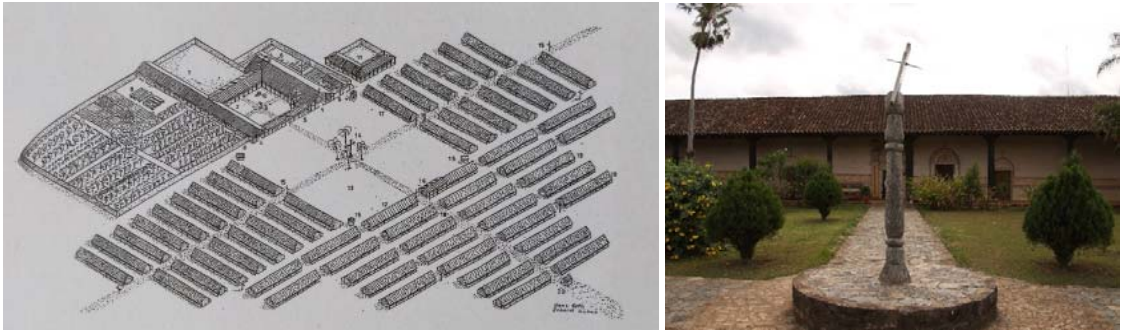

Figura 4. Plano urbano ideal de un pueblo misional de Chiquitos, inspirado en Concepción. Se detalla la avenida central – eje del pueblo – que parte de la capilla Betania (abajo a la derecha) y atraviesa la cruz de la plaza – rodeada de palmeras – para dar al patio jesuita donde se halla el reloj de Sol. Las casas de los indios cubren tres lados de la plaza; el cuarto está reservado al núcleo religioso. En la imagen, a la derecha del patio se ubica la iglesia. Dibujo de Hans Roth y Eckard Kühne, extraído de Roth (1995, 497). En la fotografía de la derecha vemos el reloj de Sol ecuatorial que hoy se halla en el patio central junto a la iglesia de San Francisco Javier. La inclinación de su tablero – donde se proyectan las sombras que marcan las horas – con respecto a la vertical es de unos 16° que corresponde a la latitud de San Javier. Así, bien orientado, el gnomon apunta al polo sur celeste, el punto imaginario alrededor del cual gira todo el cielo.

**Las iglesias de Chiquitos**

Estas iglesias son grandes pabellones sencillos que cuentan con tres naves y un amplio techo a dos aguas y de suave pendiente. Esta cobertura tradicionalmente estaba hecha con un entramado de madera donde se colocaban tejas de cerámica cocida, asentadas con barro sobre una base de cañizo. Típicamente se sustentan en dieciséis columnas de tronco macizo, procedentes de árboles de la selva tropical, talladas con un torneado que forma un diseño helicoidal –el estilo salomónico. Estas impresionantes columnas son llamadas horcones y pueden medir más de 10 metros de longitud y pesar varias toneladas. Constituyen la columna vertebral de la estructura edilicia.

En el decir del padre Cardiel (que dejó registros para las iglesias de las reducciones guaraníes, pero que es igualmente válido para las chiquitanas), "todos estos edificios se hacen de diverso modo que en Europa, porque primero se hace el tejado y después las paredes" (Furlong 1953). Esto, por supuesto, no era caprichoso, sino que respondía a la necesidad de preservar las paredes exteriores – en general de adobe y sin función resistente – de las torrenciales lluvias que azotaban estas regiones en los muy prolongados "tiempos de aguas".

Las iglesias que vemos hoy son muy amplias y – con variaciones, según cada caso particular – durante los oficios religiosos en la época jesuita podían llegar a albergar varios miles de indígenas. Ese era aproximadamente el número máximo de indios reducidos en cada pueblo misional. En aquella época contaban, además, con una profusión de elementos decorativos, de la propia arquitectura y de índole ornamental (Martini 1977, 23). En la prolongación de la nave principal se halla un presbiterio retirado, con varios arcos construidos de ladrillos, cuyas paredes laterales se encuentran en el eje de las columnas de la iglesia. De allí, a través de sendas aperturas abovedadas, se comunica con las sacristías a ambos costados de la iglesia. A los lados de las naves de la iglesia se hallan las galerías techadas y, delante de la fachada del templo, encontramos un atrio espacioso con columnas aisladas (Figura 5).

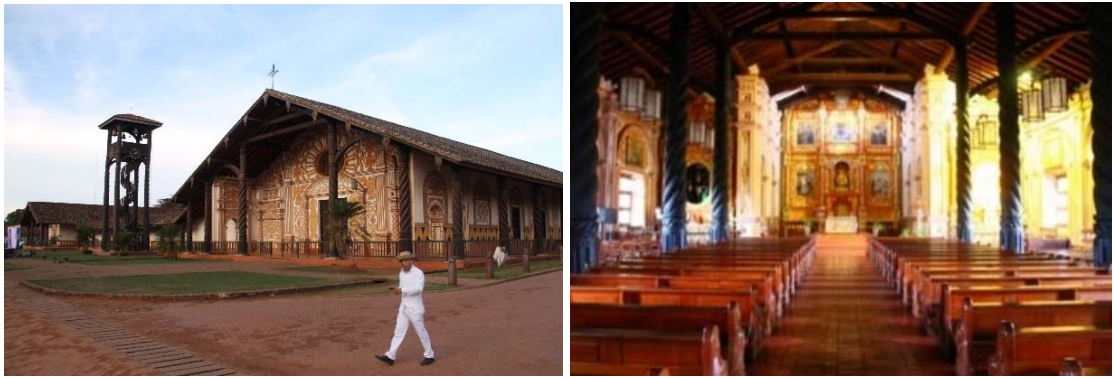

Figura 5. Iglesia Nuestra Señora de la Inmaculada Concepción. En la imagen del exterior se destaca la envergadura del templo, con el óculo en forma de Sol del frontis, el amplio atrio y la galería lateral; a la izquierda se yergue la torre-campanario. En la imagen del interior se aprecian el presbiterio y parte de las columnas de estilo salomónico que sostienen el edificio.

**Trabajo de campo**

Hemos medido la orientación espacial precisa de las iglesias con la intención de verificar la posible existencia de algún patrón definido en todo el conjunto. Este estudio con iglesias pertenecientes exclusivamente a la orden jesuita, localizadas en un territorio acotado y de difícil acceso, nos ofrece la oportunidad de verificar si las orientaciones típicas halladas en Europa se trasladaron rígidamente a esta región, o si la adaptación de los templos cristianos a estas nuevas tierras siguió otras directivas.

El trabajo consistió en la medición de la orientación astronómica (acimut y altura del horizonte local) y de la ubicación geográfica (latitud y longitud) de nueve de las diez iglesias hoy existentes. Como mencionamos, el pueblo de Santo Corazón, trasladado a su ubicación actual en fecha posterior a la expulsión de la orden, no pudo ser visitado en nuestro viaje y por ello su iglesia fue medida mediante mapas satelitales. Realizamos estas mediciones *in situ* por medio de un tándem Suunto 360PC/360R compuesto por brújula de precisión y clinómetro; estimamos que los errores están en torno a 0,5°. Para obtener la posición precisa de cada sitio se utilizó la tecnología usual de geolocalización.

La totalidad de los datos medidos y calculados para cada iglesia se hallan en la Tabla 1, donde los hemos ordenado por acimut creciente. Cada construcción en la tabla cuenta con su latitud (L) y longitud (l), el acimut (a) y el nombre de la iglesia, que refleja el santo patrón al cual está dedicada. Se incluyen, además, las fechas documentadas más aproximadas de la primera y de la última fundación de la misión y, entre corchetes, la fecha de la construcción

de la iglesia. Recordemos que la iglesia original de San Ignacio de Velasco fue demolida en 1948 y reconstruida. Además, las actuales misión e iglesia de Santo Corazón de Jesús son posteriores a la expulsión de los jesuitas. El acimut astronómico viene dado por la orientación del eje de simetría de cada construcción, medida en dirección hacia el altar y luego corregida por declinación magnética (NOAA 2020). Los valores de la declinación magnética para los diferentes sitios de las iglesias estaban comprendidos entre 12°52' y 14°39' oeste en las fechas en las que se realizaron las mediciones (octubre de 2018). En la penúltima columna de la tabla se incluye la altura angular del horizonte (h) corregida por refracción atmosférica (Schaefer 1993, 314) y, en la última columna, se consignan los valores calculados de la declinación astronómica resultante, valor que nos permite una comparación precisa con la ubicación de cualquier objeto del cielo.

| Name | L (°, S) | l (°, W) | a (°) | h (°) | δ (°) |
| --- | --- | --- | --- | --- | --- |
| San Rafael de Velasco 1696 1747 [1747] | 16.786602 | 60.674947 | 6.0 | +0.5 | +73.5 |
| Santa Ana de Velasco 1755 [c. 1760] | 16.583503 | 60.687550 | 10.0 | +0.5 | +71.2 |
| San Ignacio de Velasco 1748 [1748; new c. 1948] | 16.373513 | 60.960289 | 26.0 | 0.0 | +60.0 |
| San José de Chiquitos 1697 1706 [1725] | 17.845306 | 60.741700 | 90.5 | 0.0 | -0.6 |
| Ntra. Sra. de la Inmaculada Concepción 1708 1722 [1753] | 16.135558 | 62.023353 | 91.5 | -0.5 | -1.5 |
| San Juan Bautista (Taperas) 1699 1716 [1755] | 17.896234 | 60.376008 | 199.0 | B +1.5 | -66.3 |
| Santo Corazón de Jesús 1760 1788 [c. 1788] | 17.973958 | 58.807601 | 255.0 | B +2.5 | -15.2 |
| San Miguel de Velasco 1721 [1744] | 16.697787 | 60.968550 | 256.5 | 0.0 | -13.0 |
| San Francisco Javier 1691 1708 [1749] | 16.274553 | 62.505174 | 268.0 | 0.0 | -1.8 |
| Santiago de Chiquitos 1754 1764 [c. 1764] | 18.339631 | 59.598501 | 322.5 | +2.5 | +47.8 |

Tabla 1. Las orientaciones de las iglesias jesuíticas de la provincia de Chiquitos. Para cada construcción se consignan la latitud y longitud geográficas (L y l), el acimut astronómico (a), la altura angular del horizonte (h) detrás del altar (B significa horizonte bloqueado; se empleó el modelo de elevación digital basado en el *Shuttle Radar Topographic Mission* de Kosowsky 2017) y, en la última columna, los valores de la declinación astronómica resultantes.

**La orientación de iglesias**

En la Figura 6 mostramos el diagrama de orientación para las iglesias jesuíticas estudiadas. Los valores de los acimuts consignados son los medidos, e incluyen la corrección por declinación magnética en cada localidad. Las líneas diagonales del gráfico señalan los acimuts correspondientes – en el cuadrante oriental – a los valores extremos para el Sol (acimuts de 65,6º y 114,8º – líneas continuas –, equivalente a los solsticios de invierno y verano australes, respectivamente) y para la Luna (acimuts: 59,5º y 120,4º – líneas rayadas –, equivalente a la posición de los lunasticios mayores).

De las diez orientaciones medidas, cuatro se dirigen hacia el cuadrante norte y solo una hacia el cuadrante meridional. Las restantes cinco entran dentro del rango solar, con tres orientadas hacia poniente y dos hacia levante. Es notable que, de estas últimas cinco, hay tres que son equinocciales (con valores de declinación astronómica menores a los 2º en valor absoluto). San Francisco Javier, por ejemplo, tiene un acimut de 268º y prácticamente coincide – dentro de nuestros errores de medición – con el punto cardinal oeste. En la dirección opuesta, las iglesias de San José de Chiquitos y Nuestra Señora de la Inmaculada Concepción apuntan sus ejes hacia el este con gran precisión, la primera con un acimut de 90,5º y la segunda con 91,5º. Aunque hoy no queden registros documentales sobre las intenciones de los sacerdotes fundadores en relación con la orientación de las iglesias misionales chiquitanas, y a pesar de llegar a nuestros días un número muy exiguo de ellas como para buscar patrones en un estudio estadístico más amplio, estos resultados no dejan de ser sorprendentes.

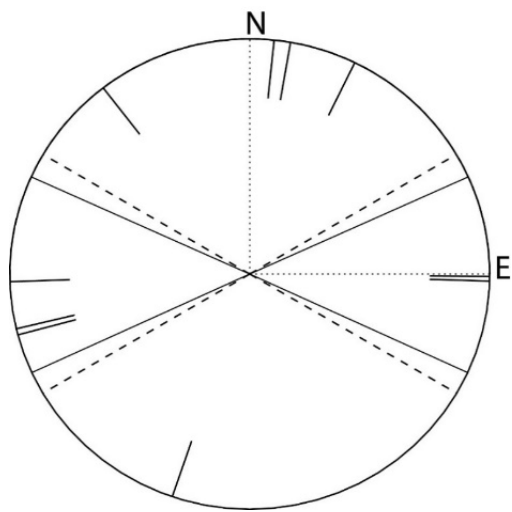

Figura 6. Diagrama de orientación para las iglesias jesuíticas de la provincia de Chiquitos, obtenido a partir de los datos de acimut de la Tabla 1.

Por otra parte, el cálculo de las declinaciones astronómicas, cuyos valores para cada iglesia se incluyen en la última columna de la Tabla 1, no presenta novedades ni sorpresas. Las alturas de los horizontes que rodean a estas iglesias tienen valores pequeños de "h". Pues en general la mayor parte de la región es de geografía llana y con suaves colinas. Además, sabemos que los jesuitas, a la hora de establecer las nuevas misiones, elegían preferentemente sitios que fueran elevados sobre el terreno local. Tenemos un ejemplo en la misión de San Miguel (fundada en 1721) que, según el padre Sánchez Labrador – quien la visitó antes de la expulsión de los jesuitas – había sido ubicada "en un buen sitio, alto y rodeado de boscaje" (Aguirre Acha 1933, 45). Esto, naturalmente, explica que los valores de "h" sean pequeños y que un gráfico de declinaciones no aporte más que lo que ya es claro a partir del diagrama de orientación de la Figura 6.

**Discusión**

Nuestros resultados indican que las orientaciones de las iglesias jesuíticas de la Chiquitanía difieren del patrón hallado para las más numerosas iglesias guaraníes (unos 30 templos documentados en el último período) que la misma orden erigió en la provincia histórica del Paraguay. Investigaciones previas (Giménez Benítez et al. 2018; ver también Ruiz Martínez-Cañavate 2017) mostraron que una gran mayoría de las 27 ruinas de las iglesias jesuíticas que pudieron ser localizadas en el noreste de la Argentina, en el sur de Paraguay y de Brasil, dirige sus ejes hacia el cuadrante meridional.

El diagrama de orientación de la Figura 6 muestra una distribución de acimuts menos concentrada hacia un cuadrante particular, aunque también permite ver que existen ciertas peculiaridades que vale la pena resaltar. Por ejemplo, solo una décima parte de las iglesias guaraníes analizadas se orientaba en el rango solar, ya sea a poniente como a levante (Giménez Benítez et al. 2018), mientras que en el caso de las chiquitanas esa proporción se eleva a la mitad de las diez analizadas.

Curiosamente, la iglesia de la Inmaculada Concepción (o mejor, sus ruinas), localizada en la mesopotamia Argentina, es la única de entre las 30 iglesias jesuíticas guaraníes originales que apunta su altar hacia el cuadrante oriental, muy cerca del este geográfico (Giménez Benítez et al. 2018). Como podemos ver en la Tabla 1, la iglesia homónima en el pueblo chiquitano de Concepción es una de las dos únicas orientadas hacia el este con gran precisión, con un acimut de 91,5º. Quizás esta no sea una mera coincidencia, pues los pueblos misionales de la Chiquitanía comenzaron a surgir muchas décadas después de establecidas las misiones de la provincia de Paraquaria (c. 1632), y es bien conocida la relación y el intercambio de documentos entre los padres de las misiones sudamericanas. Esto pudo llevar al suizo Martin Schmid, artífice de la iglesia chiquitana de Concepción, a construirla con igual orientación que su hermana guaraní.

Por otra parte, la organización de las misiones y la ubicación de los diferentes edificios parecía ser dinámica, al menos en sus primeros años. Y es muy probable que, llegado el momento de apuntalar la iglesia definitiva, el pueblo ya delineado sufriese algunas variaciones urbanísticas. Uno de los protagonistas de la reconstrucción moderna de varios templos, el arquitecto jesuita Hans Roth (Limpias Ortiz 2000), sugiere que el padre Schmid, al orientar sus iglesias hacia algún punto cardinal "con la precisión de un buen matemático y relojero suizo", debió afrontar luego el gran costo que implicó reorganizar las calles de los pueblos ya existentes para que siguieran la dirección impuesta por los ejes de las nuevas iglesias (Roth 1995, 484). Quizá no sea casualidad, entonces, que las cuatro iglesias construidas por Schmid estén todas orientadas hacia uno de los cuatro puntos cardinales: San Francisco Javier hacia el oeste, la Inmaculada Concepción, como ya vimos, hacia el este, ambas con gran precisión, y San Rafael apuntando a solo 6º del norte. El eje de San Juan Bautista, por su parte, dista del punto cardinal sur a casi 20º, pero debemos tener en cuenta que de esa iglesia quedan las ruinas solo parcialmente excavadas por los arqueólogos (en particular, su torre, quizá la construcción original no restaurada más antigua existente hoy, que data de 1755), y por ello nuestras mediciones en este caso arrastran un mayor error.

Las iglesias de San José de Chiquitos y de San Miguel de Velasco se orientan ambas dentro del rango solar y, como ya señalamos, la primera es equinoccial con muy alta precisión (acimut de 90,5º). San Ignacio de Velasco, en cambio, está orientada hacia el cuadrante norte (acimut de 26º), muy lejos del rango solar. Como es el caso de las demás iglesias, las razones de esta orientación no quedaron escritas en las crónicas de la época. Por otra parte, San Ignacio corrió una suerte muy diferente de las otras iglesias chiquitanas pues, debido a la falta de mantenimiento y a un prolongado abandono a lo largo de décadas, tuvo que ser derribada finalmente en 1948. Fue una suerte lamentable pues, según Roth (1995, 466), era "la más bella de su estilo en el interior de América". Aunque esta apreciación puede ser subjetiva, lo que sí es cierto es que, de acuerdo con los registros de la época, era el templo más grande, elaborado y adornado de la región y, quizá por eso mismo, era el de más complicada construcción. Unos años más tarde, en el mismo emplazamiento de la iglesia derribada, se construyó la nueva, que es la que vemos hoy.

Por su parte, las misiones de Santiago, Santa Ana y Santo Corazón, fundadas originariamente en el último período de los jesuitas en Chiquitos, entre los años 1750-1766, surgieron de la necesidad de dar cabida a un gran número de "infieles" que deseaba reducirse. Entre estas, Santo Corazón tuvo que trasladarse en tiempos posteriores a la expulsión de 1767, por lo cual la construcción de la iglesia actual, aún dentro del estilo jesuítico, no fue supervisada por miembros de la Orden. Notemos, además, que las iglesias de las dos primeras misiones, entre estas tres "tardías", tienen sus ejes apuntando hacia el cuadrante norte – en el caso de Santa Anta, con un acimut incluso menor que el de San Ignacio – y que se hallan bien alejadas del rango solar. Quizá la verdadera razón esté en otra parte, aunque muy probablemente no sea casualidad que, apremiados por la urgencia de albergar a tantos indígenas, la orientación de los templos – si en efecto fue relevante en construcciones previas – haya sido dejada de lado en estas circunstancias. En particular, la misión de Santa Ana muestra un claro eje regulador del pueblo que comienza en la capilla Betania, pasa luego por la cruz de la plaza y finalmente por el cuadrante y parte del colegio, pero difiere en algunos grados del eje de su iglesia (Suárez Salas 1995, 431; Roth 1995, 488). Esto puede comprobarse tanto a partir de los mapas de la misión como mediante las imágenes satelitales, pues Santa Ana, entre los pueblos menos intervenidos, conserva muchas de sus estructuras y distribución de espacios originales.

Debemos concluir entonces que, muy probablemente, las orientaciones de los pueblos misionales de Chiquitos (y, por añadidura, de las iglesias) no sigue una prescripción o un patrón definidos, pues estas debían adaptarse a cada sitio individual, lo que acarreaba no pocas dificultades. Sin embargo, a partir de nuestras mediciones hemos podido ver que existen ciertas características notorias en estas orientaciones que ha valido la pena destacar.

Para concluir, mencionemos ahora un tema que creemos que justifica un estudio futuro detallado, y que también podría estar relacionado con la orientación de las iglesias. Nos referimos a los efectos de iluminación dentro de los templos, es decir al empleo de la luz natural (el ingreso de la luz solar) para resaltar y acentuar características particulares de la arquitectura interior, por ejemplo, en el altar (ver el estudio de Vilas-Estévez y González-García (2016) para la catedral de Santiago de Compostela). Las iglesias chiquitanas tienen sus ventanas bajas en los muros laterales, a la altura de las personas, y bien protegidas por los corredores anchos de las galerías techadas. Esto hace – y hacía en su momento – que sean de por sí relativamente oscuras.

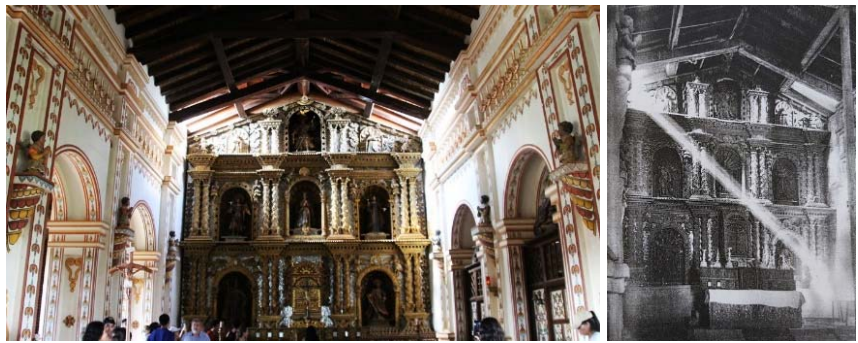

Figura 7. Lucernarios de San Rafael. Orientada hacia el cuadrante norte ($6^0$ de acimut), el Sol tropical del mediodía ya no ingresa por las ventanas laterales ni por la puerta principal de la iglesia. Los lucernarios permiten iluminar el altar con luz natural. A la derecha, imagen en blanco y negro de un haz de luz vespertino que atraviesa uno de los lucernarios del presbiterio e ilumina el retablo mayor (fotografía de 1951 de Hans Erlt; Querejazu 1995, 111).

Sin embargo, es bien conocida la relación inseparable entre la arquitectura barroca y la luz natural. Una manifestación de los efectos de luz buscados persiste aún hoy en los vistosos *lucernarios*, una suerte de claraboyas o tragaluces ubicados en los tejados que permiten iluminar los presbiterios de los templos, por ejemplo, los de San Miguel y San Rafael (Figura 7). En el segundo caso, Roth señala: "En verano, desde noviembre hasta febrero, un rayo del sol cruza lentamente el retablo mayor de San Rafael, entrando por los dos lucernarios en el techo" (Roth 1995, 507). A pesar de la ausencia de crónicas históricas jesuitas, es muy probable que los padres constructores de las iglesias tuviesen en cuenta el recorrido diurno del Sol en diferentes momentos del año. Esto les habría permitido aprovechar su luz, ya fuese para iluminar el presbiterio o para crear otros efectos notables.

Además, es conocida la veneración de los indios chiquitos por los astros más prominentes, la Luna y el Sol, y su representación sincrética se hallaba presente en los retablos de las iglesias, ya sea en forma de esculturas o de vistosas pinturas (Roth 1995, 507). También hay evidencias de que el Sol fue aprovechado para generar una atmósfera o clímax especial durante la liturgia. Por ejemplo, especialmente cerca de los equinoccios, en iglesias orientadas hacia levante como la Inmaculada Concepción y San José de Chiquitos, se producía un efecto de maravilla cuando, durante la misa de la tarde, el Sol poniente ingresaba en el templo y bañaba con su tonalidad dorada las policromías del retablo mayor y la figura de la santa o del santo patrón. Algo similar ocurría, por supuesto, en la iglesia de San Francisco Javier (y, en otras épocas del año, también en San Miguel y en Santo Corazón) pero durante el oficio matinal.

Para dar un ejemplo concreto de este efecto de luz "buscado", al menos por parte de algunos de los padres arquitectos de Chiquitos, notemos que de las diez iglesias estudiadas hay tres: San Francisco Javier, Inmaculada Concepción y San Rafael, todas construidas por el padre Schmid, que tienen un "óculo" prominente – una ventana alta con forma de Sol – en sus frontispicios (Figuras 5 y 8). Detalles decorativos aparte, sin duda esas aberturas habían sido concebidas para aprovechar al máximo el ingreso de la luz solar en ocasión de los servicios religiosos.

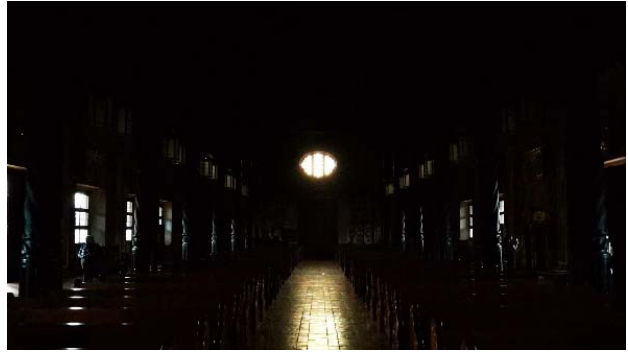

Figura 8. El óculo en forma de Sol del frontis de la iglesia Nuestra Señora de la Inmaculada Concepción visto desde el interior que permite la entrada de luz natural.

Existen estudios detallados de iluminación en muchas iglesias y catedrales alrededor del planeta (e.g., Heilbron 1999, 93, en el contexto de "meridiane", y el ya citado trabajo de Vilas-Estévez y González-García 2016), pero solo comentarios dispersos sobre este tema en el caso de los templos chiquitanos. Es un tema que sin duda requiere más investigación.

**Agradecimientos**